\documentclass[pra,showpacs,amsfonts,amsmath,floatfix,onecolumn]{revtex4} 

\usepackage{graphicx}
\usepackage{epstopdf}
\usepackage{amssymb}
\usepackage{amsmath}
\usepackage{xspace}
\usepackage{color}

\begin{document}

\title{Propagation and Distribution of Quantum Correlations in a Cavity QED Network}
\author{Raul Coto Cabrera}
\affiliation{Departamento de F\'{i}sica, Pontificia Universidad Cat\'{o}lica de Chile, Casilla 306, Santiago, Chile}
\email{rcoto@uc.cl, morszag@fis.puc.cl}
\author{Miguel Orszag}
\affiliation{Departamento de F\'{i}sica, Pontificia Universidad Cat\'{o}lica de Chile, Casilla 306, Santiago, Chile}
\email{morszag@fis.puc.cl}

\begin{abstract}
We study the propagation and distribution of quantum correlations through two chains of atoms inside cavities joined by optical fibers. We consider an effective Hamiltonian for the system and cavity losses, in the dressed atom picture, using the Generalized Master Equation. \end{abstract}


\maketitle

\section{Introduction}

Quantum Correlations rank among  the most striking features of quantum many-body systems. Entanglement is one kind of quantum correlations, which has been extensively researched, especially for a two-body system, leading to powerful applications \cite{nielsen}.

Over the past few years, the manipulation and generation of bi-partite entangled states \cite{braun,maritza}, have been widely investigated in various quantum systems such as cavity quantum electrodynamics (Cavity QED) \cite{plenio,vitalie}, trapped ions \cite{monroe}, Hubbard Model \cite{luis} and so on. For a general review, see \cite{maritza2}. However, there are other kinds of quantum correlations, such as the quantum discord \cite{QD}, which also can be responsible for computational speedup for certain quantum tasks \cite{caves}. Quantum discord has been defined as a mismatch between two quantum analogs of classically equivalent expressions of the mutual information. This notion of quantum discord goes beyond entanglement. For example, separable states can have non zero discord.
Any realistic quantum system will inevitably interact with the surrounding environment causing the rapid destruction of quantum properties. It has been observed that the quantum discord is more robust than the entanglement, against decoherence \cite{serra,boas,caldeira3}. Even in the cases where entanglement suddenly disappears, quantum discord decays only asymptotically in time \cite{Eberly}.
 
Recently, there has been a growing interest in studying atomic system in cavity QED, as well as cavity-atom polaritonic excitations \cite{angelakis1,angelakis2}.
Furthermore, much attention has been paid to the possibility of quantum information processing realized via optical fibers and more generally in schemes which allow for reliable transfer of quantum information between two atoms in distant coupled cavities \cite{jc2,pelli,victor,mancini,liang}. In the present work, we study the propagation and distribution of quantum correlations in a cavity array. We model the losses of our system with individual reservoirs at zero temperature and use the generalized master equation for the time evolution \cite{jc2,petruccione}.

\section{The Model}

\begin{figure}[ht]
\centering
\includegraphics[width=8.3cm]{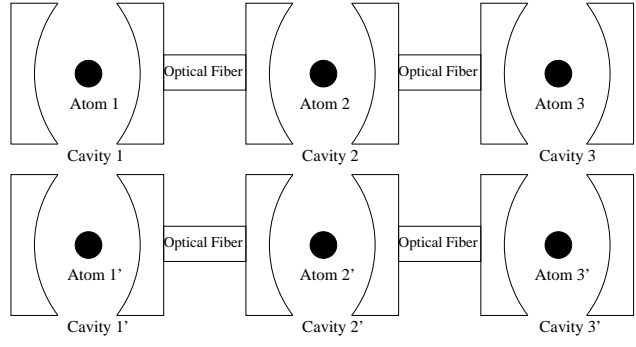}
\caption{Array of two rows of three Cavity-Atom Systems.}
\label{fig1}
\end{figure}

We have two identical chains of three cavities joined by optical fibers as shown in Fig.\ref{fig1}, where each cavity interacts with a single atom and it's own reservoir. A similar model was used by Zhang \textit{et al.} \cite{zhang} but without losses. We model our system in short fiber limit $2l\mu/2\pi c\ll 1$, where $l$ is the length of the fiber and $\mu$ is the decay rate of the cavity fields into a continuum of fiber modes \cite{serafini}.

\subsection{The Effective Hamiltonian}

 The Hamiltonian of an $N$-atom-cavity system in the rotating wave approximation is given by

\begin{equation}
\mathit{H}=\mathit{H}^{\textit{free}}+\mathit{H}^{\textit{int}}
\end{equation}

where

\begin{equation}
\mathit{H}^{\textit{free}}=\sum_{i=1}^N \omega_i^a|e>_i<e| + \sum_{i=1}^N \omega_i^c a_i^{\dagger}a_i + \sum_{i=1}^{N-1}\omega_i^{f}b_i^{\dagger}b_i
\end{equation}

and

\begin{equation}
\mathit{H}^{\textit{int}}=\sum_{i=1}^N \nu_i(a_i^{\dagger}|g>_i<e|+a_i|e>_i<g|) + \sum_{i=1}^{N-1}J_i[(a_i^{\dagger}+a_{i+1}^{\dagger})b_i+(a_i+a_{i+1})b_i^{\dagger}]
\end{equation}

where $|g>_i$ and $|e>_i$ are the ground and excited states of the two-level atom with transition frequency $\omega^a$, and $a_i^{\dagger}$($a_i$) and $b_i^{\dagger}$($b_i$) are the creation(annihilation) operators of the cavity and fiber mode, respectively. The first, second and third term in $\mathit{H}^{\textit{free}}$ are the free Hamiltonian of the atom, cavity field and fiber field, respectively. In addition, the first term in the $\mathit{H}^{\textit{int}}$ describes the interaction between the cavity mode and the atom inside the cavity with the coupling strength $\nu_i$,and the second term is the interaction between the cavity and the fiber modes with the coupling
strength $J_i$.\\

The first two terms of $\mathit{H}^{\textit{free}}$ and the first term of $\mathit{H}^{\textit{int}}$ can be jointly diagonalized in the basis of polaritons. For simplicity we consider the resonance between atom and cavity $\omega^a=\omega^c=\omega$, and also that the cavities and the fibers are identical. The total Hamiltonian is now given by

\begin{equation}
\mathit{H}=\sum_{i=1}^N(\omega-\nu)|E>_i<E|+\sum_{i=1}^{N-1}\frac{J}{\sqrt{2}}[(L_i^{\dagger}+L_{i+1}^{\dagger})b_i + (L_i^-+L_{i+1}^-)b_i^{\dagger}]
\end{equation}

where $|E_i>=\frac{1}{\sqrt{2}}(|1,g>_i-|0,e>_i)$ and $|G_i>=|0,g>_i$ are the polaritonic states, corresponding to excited and ground state respectively. The other operators $\mathit{L^\dagger_i}=|E_i><G_i|$ and $\mathit{L^-_i}=|G_i><E_i|$ are to create or destroy those states. So we can consider polaritons as a two-level system. We just can have one photon, at most, because due to photon blockade, double or higher occupancy of the polaritonic states is prohibited \cite{blockade1,blockade2}.\\

In the case of a three atom-cavity system, we use perturbation theory \cite{cohen} to find an effective Hamiltonian, supposing that the total detuning $\delta=(\omega-\nu)-\omega^f\gg J$. Furthermore, we trace over the fibers, so we end up with a reduced Hamiltonian given by:

\begin{equation}
\mathit{H_s}=\lambda(|E_1><E_1|+2|E_2><E_2|+|E_3><E_3|)+\lambda(\mathit{L^\dagger_1 \mathit{L^-_2}}+\mathit{L^-_1 \mathit{L^\dagger_2}}+\mathit{L^\dagger_2 \mathit{L^-_3}}+\mathit{L^-_2 \mathit{L^\dagger_3}})
\end{equation}  

where $\lambda=\frac{J^2}{2\delta}$.\\
\\
\subsection{The Master Equation}

Until now, we have not considered losses. The main source of dissipation originates from the leakage of the cavity photons due to imperfect reflectivity of the cavity mirrors. A second source of dissipation corresponds to atomic spontaneous emission, that we will neglect assuming long atomic lifetimes.

An approach to model the above mentioned losses, in the presence of single mode quantized  cavity fields, is using the microscopic master equation (\ref{em}), which goes back to the ideas of Davies on how to describe the system-reservoir interactions in a Markovian master equation \cite{davies}. For a three-cavity-system,

\begin{equation}\label{em}
\dot{\rho}(t)=-i\left[\mathit{H_s},\rho(t)\right]+\sum_{n=1}^3\sum_{\omega=-\infty}^{\infty}\gamma_n(\omega)\left( \mathit{A_n}(\omega)\rho(t)\mathit{{A}^\dagger_n}(\omega)-\frac{1}{2}\left\lbrace \mathit{{A}^\dagger_n}(\omega)\mathit{A_n}(\omega),\rho(t) \right\rbrace\right)   
\end{equation}

where $\mathit{A_n}$ correspond to the Davies's operators. The sum on $n$ is over all the dissipation channels and the decay rate $\gamma_n(\omega)$ is the Fourier transform of the correlation functions of the environment \cite{petruccione}.

\begin{equation}
\gamma(\omega)=\int_{-\infty}^{\infty}e^{i\omega t}<\hat{E}^{\dagger}(t)\hat{E}(0)>\, dt
\end{equation}

where the environment operators $\hat{E}$ are in the interaction picture. In our analysis, we consider $\gamma(\omega)$ as a constant \cite{victor}, and we have chosen realistic values from experiments \cite{serafini}.

The sum in equation (\ref{em}) is over positive and negative values. The corresponding damping constants are related by \cite{petruccione}, 

\begin{equation}\label{gama}
\gamma(-|\omega|)=exp(-\frac{|\omega|}{k_BT})\gamma(|\omega|)
\end{equation}

where $k_B$ is the Boltzmann's constant. Since in this work, our system is at zero temperature, it is easy to see from (\ref{gama}) that $\gamma(-|\omega|)$ vanishes, so our master equation is reduced to:

\begin{equation}\label{em2}
\dot{\rho}(t)=-i\left[\mathit{H_s},\rho(t)\right]+\sum_{n=1}^3\sum_{\omega>0}^{\infty}\gamma_n(\omega)\left( \mathit{A_n}(\omega)\rho(t)\mathit{{A}^\dagger_n}(\omega)-\frac{1}{2}\left\lbrace \mathit{{A}^\dagger_n}(\omega)\mathit{A_n}(\omega),\rho(t) \right\rbrace\right)   
\end{equation}

The $\mathit{A_n}$ operators are calculated as follows:

\begin{equation}\label{operatorA}
\mathit{A_n}(\omega_{\alpha\beta})=|\phi>_{\alpha}<\phi|(a_n + a_n^{\dagger})|\phi>_{\beta}<\phi|
\end{equation}

where $|\phi>_{\alpha}$ are the eigenstates of the Hamiltonian $\mathit{H_s}$, with $\lambda_{\alpha}$ their eigenvalues, and $\omega_{\alpha\beta}=\lambda_{\beta}-\lambda_{\alpha}$.

Since we are working at zero temperature, all transition are downwards and $\mathit{A_n}(\omega_{\alpha\beta})$ reduced to:

\begin{equation}
\mathit{A_n}(\omega_{\alpha\beta})=|\phi>_{\alpha}<\phi|a_n|\phi>_{\beta}<\phi|
\end{equation}

\section{Numerical Results}

We consider the six-cavity-system of Fig.$1$ and use the following notation, $|\Psi>=|X_1X_{1'}X_2X_{2'}X_3X_{3'}>$, where $X$ could be $G$ or $E$. We studied two different initial conditions with only excitations in cavities $1$ and $1'$:
 
\begin{eqnarray}
&|\Psi>_a=\sin(\theta)|GEGGGG> + \cos(\theta)|EGGGGG>\nonumber\\
&|\Psi>_b=\sin(\theta)|GGGGGG> + \cos(\theta)|EEGGGG>
\end{eqnarray}

\subsection{Propagation of Entanglement}

We studied the evolution of the \textit{Concurrence} \cite{wootters}, in time,  its distribution over the system, and the way it propagates.

If we set $\rho_{AB}$ to be the density matrix of a two-qubit system $A$ and $B$, then, we define the ``spin-flipped'' density matrix

\begin{equation}
\tilde{\rho}_{AB}=(\sigma_y\otimes\sigma_y)\rho_{AB}^{\ast}(\sigma_y\otimes\sigma_y)   
\end{equation}
 
 where $\sigma_y$ is the usual Pauli's matrix. Then the concurrence of the density matrix $\rho_{AB}$ is defined as 
 
 \begin{equation}
 \mathit{C}_{AB}=\max{\lbrace0,\alpha_1-\alpha_2-\alpha_3-\alpha_4\rbrace}
 \end{equation} 

where the $\alpha_1,\,\alpha_2,\,\alpha_3,\,\alpha_4$ are the square root of the eigenvalues of $\rho_{AB}\tilde{\rho}_{AB}$ in decreasing order.

We now turn to our first problem, the propagation, with the following parameters, $J=2\pi\cdot30\,GHz$, $\delta=2\pi\cdot300\,GHz$ and $\gamma=0.01\, GHz$.

We found that the transmission properties of the entanglement depend strongly on the initial conditions. For example, we observed that for the initial state $|\Psi>_a$, $74.2\%$ of the concurrence in the cavity-pair $11'$ is transmitted to the $33'$ pair, independent of the angle $\theta$. On the other hand, for $|\Psi>_b$ the transmission depends strongly on $\theta$. For example, for $\theta=\pi/3$ we get $63\%$ and for $\theta=\pi/8$, $28\%$.

\begin{figure}[ht]
\centering
\includegraphics[height=5cm,width=8cm]{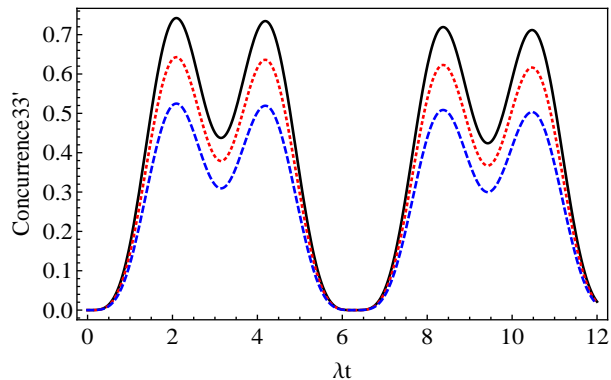}
\caption{ Concurrence for the initial condition $|\Psi>_a$;\hspace*{0.1cm}$\theta =\pi/4$(solid);$\theta =\pi/3$(red-dotted);$\theta =\pi/8$(blue-dashed);$\gamma=0.01$; for the cavities $33'$.}
\label{fig2}
\end{figure}

The final concurrence $33'$ is shown in Fig.\ref{fig2}, for the initial state $|\Psi>_a$.\\

According to Wootters \textit{et al.} \cite{wootters2}, for pure states, one can define $\mathit{C}^2_{i(jk...)}=4\det\rho_i$, where $\rho_i$ is the reduced density matrix, which represents the square of the concurrence between the cavity $``i"$ and the rest. Therefore, rather than plotting the time behavior of the concurrences of all the possible combinations of the $6$ cavities, we prefer to study the time evolution of $\mathit{C}^2_{i(jk...)}$. \\

\begin{figure}[ht]
\centering
\includegraphics[height=5cm,width=8cm]{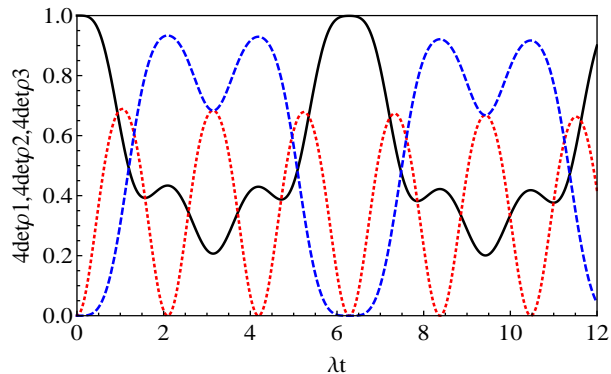}
\caption{Initial condition $|\Psi>_a$;\hspace*{0.2cm}$4\det\rho_1$(solid);$4\det\rho_2$(red-dotted);$4\det\rho_3$(blue-dashed);$\theta=\pi/4$;$\gamma=0$.}
\label{fig3}
\end{figure}

In Fig.\ref{fig3} we show such a behavior. For example, we notice that at a particular time, $4\det\rho_2=0$, implying that all the concurrences involving cavity $2$ vanish.

\begin{figure}[ht]
\centering
\includegraphics[height=1.5cm,width=9cm]{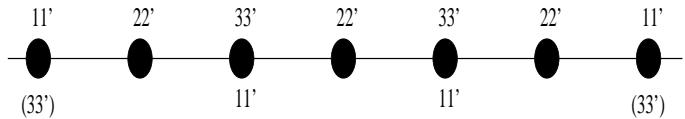}
\caption{Scheme of the time evolution of the concurrence.}
\label{fig4}
\end{figure}

An alternative way of describing the propagation of the entanglement is shown in Fig.\ref{fig4}, where we follow the appearance of the maximum values of the concurrences between the cavities $11',22'$ and  $33'$.

The main idea behind this scheme is that we can picture the propagation of the concurrence for any initial Bell state between the cavities $11',13'$ or any of the equivalent combinations. That is because we can rotate the second chain and still have the same problem, without actually solving the master equation in each case. 
For example, if we start with the $11'$ combination, the dynamics will follow the sequence in Fig.\ref{fig4}, excluding the terms in parenthesis; and the points that show two pairs of numbers imply that we have two simultaneous and in general different maxima in the concurrence. 
On the other hand, if we switch to the initial condition  for the $22'$ pair, we observe the same time sequence as in the previous case, except for the terms in parenthesis, that indicate simultaneous maxima in $11'$ and $33'$. Furthermore, for that initial condition ($22'$), we always have $\mathit{C}_{11'}=\mathit{C}_{33'}$.  


\subsection{Propagation of Quantum Discord}

In this subsection we investigate the dynamics of the quantum discord, which is the difference between the quantum mutual information and the classical correlation.

\begin{equation}
\mathit{Q}(\rho_{AB})=\mathit{I}(\rho_{AB})-\mathit{CC}(\rho_{AB})
\end{equation}

The mutual information  $\mathit{I}(\rho_{AB})$ of two subsystem can be expressed as 

\begin{equation}
\mathit{I}(\rho_{AB})=S(\rho_A)+S(\rho_B)-S(\rho_{AB})
\end{equation}

where $S(\rho)=-tr(\rho\log_2\rho)$ is the von Neumann entropy, and $\rho_A$($\rho_B$) is the reduced density matrix of subsystem $A$($B$).

The classical correlation $\mathit{CC}(\rho_{AB})$ is defined as the maximum information that one can obtain from $A$ by performing a measurement on $B$, and in general this definition is not symmetric.

\begin{equation}\label{CC}
\mathit{CC}(\rho_{AB})=\max_{\lbrace B_k \rbrace}[S(\rho_A)-S(\rho_{AB}|\lbrace B_k\rbrace)]
\end{equation}    

where $\lbrace B_k\rbrace$ is a complete set of projectors performed on subsystem $B$ and $S(\rho_{AB}|\lbrace B_k\rbrace)=\sum_k p_k S(\rho_A^k)$. The reduced density operator $\rho^k$ associated with the measurement result $k$ is:

\begin{equation}
\rho^k=\dfrac{1}{p_k}(\mathit{I}\otimes B_k)\rho(\mathit{I}\otimes B_k)
\end{equation}

Notice that the probability $p_k$ can be easily obtained by taking the trace over the last equation. Instead of finding the maximum in (\ref{CC}), we will minimize the second term in the same equation, which is equivalent. There are different ways of doing that \cite{luo,alber,caldeira3}, and for simplicity we will follow the one in reference \cite{alber}. 

In some of the following calculations, when comparing the various measures of correlations,  it will be  more convenient to calculate the entanglement of formation ($E$), rather than the concurrence.
The connection between the two is given by a simple formula \cite{wootters} 

\begin{eqnarray}
E(\mathit{C})&=&h(\dfrac{1+\sqrt{1-\mathit{C}^2}}{2})\nonumber\\
h(x)&=&-x\log_2(x)-(1-x)\log_2(1-x)
\end{eqnarray}

Next, we plot both, the quantum discord and the entanglement of formation for the cavity $3$ and $3'$, and we study the time evolution of the system.

\begin{figure}[ht]
\centering
\includegraphics[height=5cm,width=8cm]{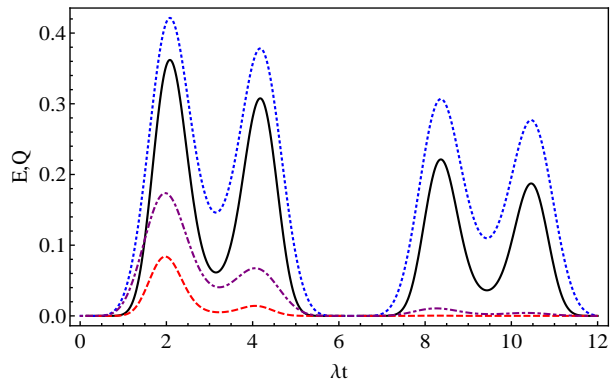}
\caption{Initial condition $|\Psi>_b$;\hspace*{0.2cm} $E$(solid); $Q$(blue-dotted); for $\gamma=0.05$;\hspace*{0.2cm}$E$(red-dashed); $Q$(purple-dot dashed); for $\gamma=0.5$;\hspace*{0.2cm}$\theta=\pi/4$; for the cavities $33'$.}
\label{fig5}
\end{figure}

From Fig.\ref{fig5} we can see that the quantum discord, remains bigger than the entanglement of formation. Notice that $\mathit{Q}$ shows up first and it holds different from zero for a longer time than $E$. Asymptotically, for all $\gamma$, $\mathit{Q}$ tends to be above $E$, even when the latter vanishes, which is in agreement with previous work in cavity QED \cite{bellomo}. On the other hand, we observe twin peaks that appear periodically with decreasing amplitudes. This behavior is consistent with the curve for $4\det\rho_3$ shown in the Figure (3).

We are also interested in the classical correlations . If we take the initial condition $|\Psi_b>$, the quantum and classical correlations between the cavity pairs $11'$,$22'$ and $33'$ are identical for all times.
However, for any other initial condition for the $11'$ pair, such as the mixed state $\rho(0)$

\begin{equation}
\rho(0)=\dfrac{1}{2}|EE><EE| + \dfrac{1}{2}|GG>GG| + \dfrac{1}{2}(|EE><GG|+|GG><EE|),
\end{equation}
the various measures of quantum correlations are all different, as shown in Fig.(6), when taking, for example, the time evolution of the correlations between the $21'$ pair of cavities.
Furthermore, for relatively small damping constants, we observe for some time intervals, that both the quantum and classical correlations remain approximately constant (See Fig.(6)).

\begin{figure}[ht]
\centering
\includegraphics[height=5cm,width=8cm]{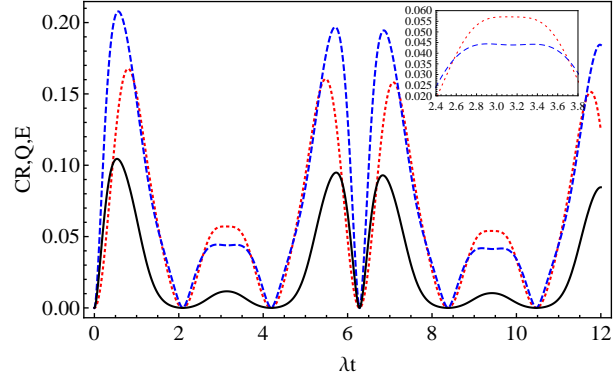}
\caption{Initial condition $\rho(0)$;\hspace*{0.1cm} $\mathit{CC}$(red-dotted);$Q$(blue-dashed);$E$(solid);\hspace*{0.1cm}$\gamma=0.01$; for the cavities $21'$.}
\label{fig6}
\end{figure}
\subsection{Distribution of Entanglement}

Coffman \textit{et al.} \cite{wootters2} discuss distributed entanglement \cite{buzek1,buzek2}. They argue that unlike classical correlations, quantum entanglement cannot be freely shared among many objects.
In the case of three partite system,$S_1,\,S_2$ and $S_3$; the amount of entanglement that $S_1$ can share with $S_2$ and $S_3$, must satisfy an inequality
 
\begin{equation}\label{relation}
\mathit{C}_{12}^2 + \mathit{C}_{13}^2\leq 4\det\rho_1
\end{equation}

with $\rho_1=tr_{23}\rho_{123}$.\\

More recently, the trade off between entanglement and classical correlation  has been investigated and a conservation law for distributed entanglement of formation and quantum discord has been found \cite{QD,winter,caldeira}. 
In the present problem, we choose the cavity $1$ as the reference one, and consider the initial pure entangled  state of the subsystems 1 and $1^{\prime}$.

It is interesting to notice that for the initial conditions $|\Psi>_a$, we get higher values of concurrence than for $|\Psi>_b$, which is in agreement with previous work. For example in reference \cite{mancini}, the authors observed  that in a qubit network with quasi local dissipation, the maximum stationary concurrence that can be achieved with an initial state containing one excitation over $m$ qubits is always higher than the state containing N excitations $(N\leq m)$, based on numerical results.  
So, in our multi atom-cavity system, we define $\mathit{S}\equiv\mathit{C}_{12}^2 + \mathit{C}_{13}^2 + \mathit{C}_{11^{\prime}}^2 +\mathit{C}_{12^{\prime}}^2+\mathit{C}_{13^{\prime}}^2$, and observed that $\mathit{S}$ is bigger for the initial state $|\Psi>_a$, with one excitation, than for $|\Psi>_b$, that contains two excitations. \\

Now, we turn to the next question: Is there a global entanglement between all the subsystems, beyond the two-partite entanglement?
To answer this question, we use the \textit{Tangle} $``\tau"$ \cite{wootters2} defined as:

\begin{equation}\label{tangle}
\tau=\mathit{C}^2_{1(23...)}-\mathit{C}_{12}^2-\mathit{C}_{13}^2-\mathit{C}_{11^{\prime}}^2 -\mathit{C}_{12^{\prime}}^2-\mathit{C}_{13^{\prime}}^2
\end{equation}
that represents the multipartite correlations (beyond 2) of the system. We can express the first term of the right side in an alternative way: $\mathit{C}^2_{1(23...)}=2(1-tr\left[ \rho^2_1\right] )$ \cite{buzek}.\\

\begin{figure}[ht]
\centering
\includegraphics[height=6cm,width=8cm]{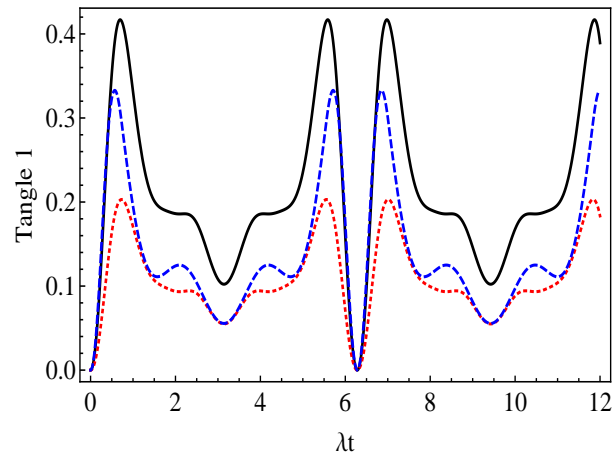}
\caption{Tangle for the initial condition $|\Psi>_b$;\hspace*{0.2cm}$\theta =\pi/4$(solid);$\theta =\pi/3$(red-dotted);$\theta =\pi/8$(blue-dashed);$\gamma=0$}
\label{fig7}
\end{figure}

In Fig.(7) we show the evolution of $\tau$ for the initial state $|\Psi>_b$ and various values of ($\theta$). In all cases, the initial tangle is zero, since we start with a bi-partite entanglement between the first pair of cavities and therefore there are no higher order correlations. 
We also found that for the initial state $|\Psi>_a$, the tangle is zero at all times. A possible reason for this is that tangle is a collective effect and thus it requires more than one excitation in the system.  \\

Up to now, our system had no losses, so the states are pure at all times. But what happens if we turn on the interaction with the reservoirs? First, the system become mixed, and equation (\ref{tangle}) is no longer correct, and we need a convex roof optimization of $\mathit{C}^2_{i(jk...)}$, considering all possible pure state decomposition of $\rho=\sum_{i}p_i|\phi>_i<\phi|$, which is given by:

\begin{equation}\label{minimization}
\mathit{C}^2_{i(jk...)}(\rho)=\inf_{\left\lbrace p_i,|\phi_i>\right\rbrace }\sum_i p_i\mathit{C}^2_{i(jk...)}(|\phi_i>)
\end{equation}

The solution of (\ref{minimization}) is a complicated task \cite{horodecki}. But there are some good approximations \cite{davidovich,upperbound,lowerbound,quasipure}. The upper bound \cite{upperbound} for this equation is taking just a pure state and the lower bound \cite{lowerbound} is taking the expression $\mathit{C}^2_{i(jk...)}(\rho)=2(Tr\left[ \rho^2\right] -Tr\left[ \rho^2_i\right] )$; where $Tr\left[ \rho^2\right]$ is the purity of the total system. Next, we study the time evolution of both bounds.

\begin{figure}[ht]
\centering
\includegraphics[height=6cm,width=8cm]{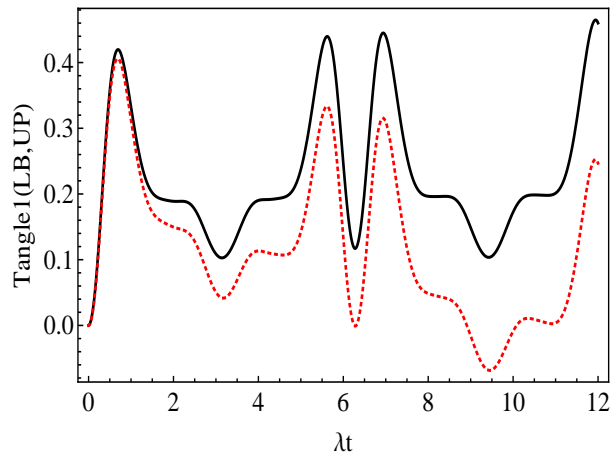}
\caption{Tangle for the initial condition $|\Psi>_b$;\hspace*{0.2cm}Upper Bound(UB)(solid);Lower Bound(LB)(red-dotted);$\theta =\pi/4$;$\gamma=0.01$}
\label{fig8}
\end{figure}

In Fig.(8) we observe the upper and lower bounds of the tangle. In the lower bound approximation, we need to guarantee that the system is weakly mixed and strongly entangled.
In particular for $\lambda t\approx 9$, the lower bound becomes negative. On the other hand, from the Fig.\ref{fig3}, we notice that in this region, $\mathit{C}^2_{1(23...)}$ is comparatively small, thus violating the assumptions made by the lower bound approximation, and therefore the results are unreliable.
Nevertheless, for $\lambda t \in\lbrace 0,6\rbrace$, the area between the upper and lower bound is rather small, giving us a good estimation of the tangle.  

\section{Summary and Conclusions}

In the present work we studied a cavity QED system with six cavities, and their corresponding atom inside, in a configuration shown in Fig.\ref{fig1}.
This type of system can be easily realized experimentally. It also can be used for various purposes, such as a channel for the propagation of quantum correlations or a network to distribute entanglement. 

If the system is used as a channel, our best option is to use states like $|\Psi>_a$ as the initial condition, since the distribution or multipartite entanglement vanishes, finding only bi-partite quantum correlations and as a consequence we get higher values of the concurrence at the extreme of the chains. For low losses, the Entanglement of formation is a good measure of the quantum correlations. However, as showed before, the Quantum Discord is more robust against decoherence, thus is a better measure for higher loss rates. 
Next, we focus on the quantum correlations between the cavities $21'$, finding time intervals where the classical and quantum correlations become approximately constants.

On the other hand, for pure states, if our purpose is to distribute the quantum correlations among the various elements of our system, we choose $|\Psi>_b$ as our initial condition, since we have a considerable multipartite entanglement or tangle. Furthermore, we observe from the Fig.(7) that the tangle deteriorates rapidly, as we depart from the Bell states ($\theta=\pi/4$). 

If we now turn on the interaction with the individual reservoirs, the situation becomes more involved, and in principle it would require a complex convex roof optimization procedure. Nevertheless, when the system experiences losses, if these are moderate, we can still estimate lower and upper bounds to the tangle, in the case where the mixedness of the system, measured through $Tr[\rho^2]$, varies slowly between $1$ and $0.89$ for $\gamma=0.01$. For higher losses, like $\gamma=0.1$, the gap between the bounds is significantly bigger and above approximations fail.

\begin{figure}[ht]
\centering
\includegraphics[height=6cm,width=8cm]{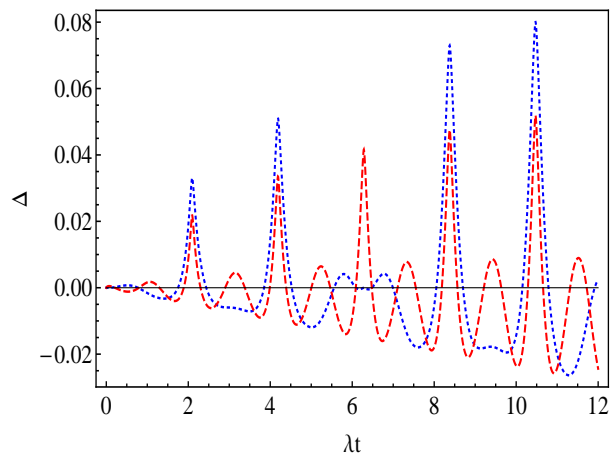}
\caption{$|\Psi>_1$(red-dashed);\hspace*{0.2cm}$|\Psi>_2$(blue-dotted);$\gamma=0.01$}
\label{fig9}
\end{figure}

Finally, we compare these results with Fanchini \textit{et al.} \cite{caldeira}, where the authors analyze a conservation law involving both entanglement of formation and quantum discord.
They find, in a three partite system, a stronger version of the ``strong subadditivity of entropy inequality''  $\mathit{S_2}+\mathit{S_3}\leq\mathit{S_{12}}+\mathit{S_{13}}$(SS)\cite{ss}, that reads:

\begin{equation}\label{mss}
\mathit{S_2}+\mathit{S_3}+\Delta \leq \mathit{S_{12}}+\mathit{S_{13}}
\end{equation} 

where $\Delta=E_{12}+E_{13}-\mathit{Q}_{12}-\mathit{Q}_{13}$. 
Of course the inequality (\ref{mss}) is stronger than (SS) provided $\Delta>0$.
In the present cavity QED system, and taking only a single chain, we plot in Fig.(\ref{fig9}) $\Delta$ versus time for two different initial conditions $|\Psi>_1=1/\sqrt{2}(|EGG>+|GGE>)$ and $|\Psi>_2=|EGG>$ and we clearly see that we have  the stronger inequality (\ref{mss}), as compared to the inequality (SS) only during short time intervals corresponding to the sharp positive peaks. However, as time goes on, it tends to drop to the negative side.
So, the stronger inequality (\ref{mss}) is valid for short times and, in general, weak interaction of the system with the environment.

\section{Acknowledgments}

M. Orszag acknowledges financial support from Fondecyt, Project $1100039$ and Programa de Investigacion Asociativa anillo ACT-1112. R. Coto thanks the support from the Pontificia Universidad Cat\'olica de Chile.  


\end{document}